\documentstyle[11pt,aaspp4,flushrt]{article}
\def \beq{\begin{equation}}
\def \eeq{\end{equation}}
\def \beqa{\begin{eqnarray}}
\def \eeqa{\end{eqnarray}}
\begin{document}

\title{The Photo-Evaporation of Dwarf Galaxies During Reionization}

\author{Rennan Barkana\footnote{email: barkana@ias.edu}}
\affil{Institute for Advanced Study, Olden Lane, Princeton,
NJ 08540}

\author{Abraham Loeb\footnote{email: aloeb@cfa.harvard.edu}}
\affil{Astronomy Department, Harvard University, 60 Garden 
St., Cambridge, MA 02138}

\begin{abstract}
During the period of reionization the Universe was filled with a
cosmological background of ionizing radiation. By that time a
significant fraction of the cosmic gas had already been incorporated
into collapsed galactic halos with virial temperatures $\la 10^4$K
that were unable to cool efficiently. We show that photoionization of
this gas by the fresh cosmic UV background boiled the gas out of the
gravitational potential wells of its host halos. We calculate the
photoionization heating of gas inside spherically symmetric dark
matter halos, and assume that gas which is heated above its virial
temperature is expelled. In popular Cold Dark Matter models, the
Press-Schechter halo abundance implies that $\sim 50$--$90\%$ of the
collapsed gas was evaporated at reionization. The gas originated from
halos below a threshold circular velocity of $\sim10$--15 km s$^{-1}$.
The resulting outflows from the dwarf galaxy population at redshifts
$z=5$--10 affected the metallicity, thermal and hydrodynamic state of
the surrounding intergalactic medium. Our results suggest that
stellar systems with a velocity dispersion $\la 10~{\rm km~s^{-1}}$,
such as globular clusters or the dwarf spheroidal galaxies of the
Local Group, did not form directly through cosmological collapse at
high redshifts.
\end{abstract}


\section{Introduction}

The formation of galaxies is one of the most important, yet unsolved,
problems in cosmology. The properties of galactic dark matter halos are
shaped by gravity alone, and have been rigorously parameterized in
hierarchical Cold Dark Matter (CDM) cosmologies (e.g., Navarro, Frenk, \&
White 1997).
However, the complex processes involving gas dynamics, chemistry and
ionization, and cooling and heating, which are responsible for the
formation of stars from the baryons inside these halos, have still not been
fully explored theoretically.

Recent theoretical investigations of early structure formation in CDM
models have led to a plausible picture of how the formation of the
first cosmic structures leads to reionization of the intergalactic
medium (IGM). The bottom-up hierarchy of CDM cosmologies implies that
the first gaseous objects to form in the Universe have a low-mass,
just above the cosmological Jeans mass of $\sim 10^4 M_\odot$ (see,
e.g., Haiman, Thoul, \& Loeb 1996, and references therein). The
virial temperature of these gas clouds is only a few hundred K, and so
their metal-poor primordial gas can cool only due to the formation of
molecular hydrogen, ${\rm H_2}$. However, ${\rm H_2}$ molecules are
fragile, and were easily photo-dissociated throughout the Universe by
trace amounts of starlight (Stecher \& Williams 1967; Haiman, Rees, \&
Loeb 1996) that were well below the level required for complete
reionization of the IGM. Following the prompt destruction of their
molecular hydrogen, the early low-mass objects maintained virialized
gaseous halos that were unable to cool or fragment into stars. Most
of the stars responsible for the reionization of the Universe formed
in more massive galaxies, with virial temperatures $T_{\rm vir}\ga
10^4$K, where cooling due to atomic transitions was possible. The
corresponding mass of these objects at $z\sim 10$ was $\sim10^8~{\rm
M_\odot}$, typical of dwarf galaxies.

The lack of a Gunn-Peterson trough and the detection of Ly$\alpha$ emission
lines from sources out to redshifts $z=5.6$ (Weymann et al.\ 1998; Dey et
al.\ 1998; Spinrad et al.\ 1998; Hu, Cowie, \& McMahon 1998) demonstrates
that reionization due to the first generation of sources must have occurred
at yet higher redshifts; otherwise, the damping wing of Ly$\alpha$
absorption by the neutral IGM would have eliminated the Ly$\alpha$ line in
the observed spectrum of these sources (Miralda-Escud\'e 1998). Popular
CDM models predict that most of the intergalactic hydrogen was ionized at a
redshift $8\la z\la 15$ (Gnedin \& Ostriker 1997; Haiman \& Loeb
1998a,c). The end of the reionization phase transition resulted in the
emergence of an intense UV background that filled the Universe and heated
the IGM to temperatures of $\sim 1$--$2\times 10^4$K (Haiman \& Loeb 1998b;
Miralda-Escud\'e, Haehnelt, \& Rees 1998). After ionizing the rarefied IGM
in the voids and filaments on large scales, the cosmic UV background
penetrated the denser regions associated with the virialized gaseous halos
of the first generation of objects. Since a major fraction of the collapsed
gas had been incorporated by that time into halos with a virial temperature
$\la 10^4$K, photoionization heating by the cosmic UV background could have
evaporated much of this gas back into the IGM. No such feedback was
possible at earlier times, since the formation of internal UV sources was
suppressed by the lack of efficient cooling inside most of these objects.

The gas reservoir of dwarf galaxies with virial temperatures $\la
10^4$K (or equivalently a 1D velocity dispersion $\la 10~{\rm
km~s^{-1}}$) could not be immediately replenished. The suppression of
dwarf galaxy formation at $z>2$ has been investigated both
analytically (Rees 1986; Efstathiou 1992) and with numerical
simulations (Thoul \& Weinberg 1996; Quinn, Katz, \& Efstathiou 1996;
Weinberg, Hernquist, \& Katz 1997; Navarro \& Steinmetz 1997). The
dwarf galaxies which were prevented from forming after reionization
could have eventually collected gas at $z=1$--2, when the UV
background flux declined sufficiently (Babul \& Rees 1992; Kepner,
Babul, \& Spergel 1997). The reverse process during the much earlier
reionization epoch has not been addressed in the literature.
(However, note that the photo-evaporation of gaseous halos was
considered by Bond, Szalay, \& Silk (1988) as a model for Ly$\alpha$
absorbers at lower redshifts $z\sim 4$.)

In this paper we focus on the reverse process by which gas that had
already settled into virialized halos by the time of reionization was
evaporated back into the IGM due to the cosmic UV background which
emerged first at that epoch. The basic ingredients of our model are
presented in \S 2. In order to ascertain the importance of a
self-shielded gas core, we include a realistic, centrally concentrated
dark halo profile and also incorporate radiative transfer. Generally
we find that self-shielding has a small effect on the total amount of
evaporated gas, since only a minor fraction of the gas halo is
contained within the central core. Our numerical results are
described in \S 3. In particular, we show the conditions in the
highest mass halo which can be disrupted at reionization. We also use
the Press-Schechter (1974) prescription for halo abundance to
calculate the fraction of gas in the Universe which undergoes the
process of photo-evaporation. Our versatile semi-analytic approach
has the advantage of being able to yield the dependence of the results
on a wide range of reionization histories and cosmological parameters.
Clearly, the final state of the gas halo depends on its dynamical
evolution during its photo-evaporation. We adopt a rough criterion for
the evaporation of gas based on its initial interaction with the
ionizing background. The precision of our results could be tested in
specific cases by future numerical simulations. In \S 4 we discuss
the potential implications of our results for the state of the IGM and
for the early history of low-mass galaxies in the local Universe.
Finally, we summarize our main conclusions in \S 5.

\section{A Model for Halos at Reionization}

We consider gas situated in a virialized dark matter halo. We adopt the
prescription for obtaining the density profiles of dark matter halos at
various redshifts from the Appendix of Navarro, Frenk, \& White (1997,
hereafter NFW), modified to include the variation of the collapse
overdensity $\Delta_c$. Thus, a halo of mass $M$ at redshift $z$ is
characterized by a virial radius, 
\beq r_{\rm vir}=0.756
\left(\frac{M}{10^8\ h^{-1} \ M_{\sun}}\right)^{1/3} \left[\frac{\Omega_0}
{\Omega(z)}\ \frac{\Delta_c}{200}\right]^{-1/3} \left
( \frac{1+z}{10}\right)^{-1}\ h^{-1}\ {\rm kpc}\ , \eeq or a corresponding
circular velocity, \beq V_c=\left(\frac{G M}{r_{\rm vir}}\right)^{1/2}=31.6
\left(\frac{r_{\rm vir}}{h^{-1}\ {\rm kpc}}\right)\left[\frac
{\Omega_0}{\Omega(z)}\ \frac{\Delta_c}{200}\right]^{1/2}
\left(\frac{1+z}{10}\right)^{3/2}\ {\rm km\ s}^{-1}\ . \eeq The density
profile of the halo is given by \beq \rho(r)=\frac{3 H_0^2}{8 \pi G}
(1+z)^3 \frac{\Omega_0}{\Omega(z)} \frac{\delta_c} {c x (1+c x)^2}\ ,
\label{NFW} 
\eeq where $x=r/r_{\rm vir}$ and $c$ depends on $\delta_c$ for a given mass
$M$. We include the dependence of halo profiles on $\Omega_0$ and
$\Omega_{\Lambda}$, the current contributions to $\Omega$ from
non-relativistic matter and a cosmological constant, respectively (see
Appendix A for complete details).

Although the NFW profile provides a good approximation to halo
profiles, there are indications that halos may actually develop a core
(e.g., Burkert 1995; Kravtsov et al.\ 1998; see, however, Moore et
al.\ 1998). In order to examine the sensitivity of the results to
model assumptions, we consider several different gas and dark matter
profiles, keeping the total gas fraction in the halo equal to the
cosmological baryon fraction. The simplest case we consider is an
equal NFW profile for the gas and the dark matter. In order to include
a core, instead of the NFW profile of equation (\ref{NFW}) we also
consider the density profile of the form fit by Burkert (1995) to
dwarf galaxies, \beq \rho(r)=\frac{3 H_0^2}{8 \pi G} (1+z)^3
\frac{\Omega_0}{\Omega(z)} \frac{\delta_c} {(1+b x) \left[1+(b
x)^2\right]}\,
\label{core} 
\eeq where $b$ is the inverse core radius, and we set $\delta_c$ by
requiring the mean overdensity to equal the appropriate value, $\Delta_c$,
in each cosmology (see Appendix A). We also consider two cases where the
dark matter follows an NFW profile but the gas is in hydrostatic
equilibrium with its density profile determined by its temperature
distribution. In one case, we assume the gas is isothermal at the halo
virial temperature, given by \beq
\label{tvir} T_{\rm vir}=\frac{\mu V_c^2}{2 k_B}=36100\ \frac{\mu}{0.6\, m_p}
\left(\frac{r_{\rm vir}}{h^{-1}\ {\rm kpc}}\right)^2 \frac
{\Omega_0}{\Omega(z)}\ \frac{\Delta_c}{200} \left(\frac{1+z}{10}\right)^3\
{\rm K} \ , \eeq where $\mu$ is the mean molecular weight as determined by
ionization equilibrium, and $m_p$ is the proton mass. The spherical
collapse simulations of Haiman, Thoul, \& Loeb (1996) find a post-shock gas
temperature of roughly twice the value given by equation (\ref{tvir}), so
we also compare with the result of setting $T=2\ T_{\rm vir}$. In the
second case, we let the gas cool for a time equal to the Hubble time at the
redshift of interest, $z$. Gas above $10^4$K cools rapidly due to atomic
cooling until it reaches a temperature near $10^4$K, where the cooling time
rapidly diverges. In this case, hydrostatic equilibrium
yields a highly compact gas cloud when the halo virial temperature is
greater than $10^4$K. In reality, of course, a fraction of the gas may
fragment and form stars in these halos. However, this caveat hardly
affects our results since only a small fraction of the gas which evaporates
is contained in halos with $T_{\rm vir}>10^4$K. Throughout most of our
subsequent discussion we consider the simple case of identical NFW 
profiles for both the dark matter and the gas, unless indicated otherwise.

We assume a helium mass fraction of $Y=0.24$, and include it in the
calculation of the ionization equilibrium state of the gas as well as its
cooling and heating (see, e.g., Katz, Weinberg, \& Hernquist 1996). We
adopt the various reaction and cooling rates from the literature, including
the rates for collisional excitation and dielectronic recombination from
Black (1981); the recombination rates from Verner \& Ferland (1996), and
the recombination cooling rates from Ferland et al.\ (1992) with a fitting
formula by Miralda-Escud\'{e} (1998, private communication). Collisional
ionization rates are adopted from Voronov (1997), with the corresponding
cooling rate for each atomic species given by its ionization rate
multiplied by its ionization potential. We also include cooling by
Bremsstrahlung emission with a Gaunt factor from Spitzer \& Hart (1971),
and by Compton scattering off the microwave background (e.g., Shapiro \&
Kang 1987).

In assessing the effect of reionization, we assume for simplicity a sudden
turn-on of an external radiation field with a specific intensity per unit
frequency, $\nu$, \beq I_{\nu,0}=10^{-21}\, I_{21}(z)\, (\nu/\nu_L)^
{-\alpha}\mbox{ erg cm}^{-2}\mbox{ s}^{-1}\mbox{ sr}
^{-1}\mbox{Hz}^{-1}\ , \label{Inu} \eeq 
where $\nu_L$ is the Lyman limit frequency. Our
treatment of the response of the cloud to this radiation, as outlined
below, is not expected to yield different results with a more gradual
increase of the intensity with cosmic time. The external intensity
$I_{21}(z)$ is responsible for the reionization of the IGM, and so we
normalize it to have a fixed number of ionizing photons per baryon in the
Universe. We define the ionizing photon density as \beq
n_{\gamma}=\int_{\nu_L}^{\infty} \frac{4 \pi I_{\nu,0}}{h \nu c}\, \frac
{\sigma_{HI}(\nu)}{\sigma_{HI}(\nu_L)}\,d\nu \ ,
\label{eq:n_gamma}
\eeq where the photoionization efficiency is weighted by the
photoionization cross section of $HI$, $\sigma_{HI}(\nu)$, above the
Lyman limit. The mean baryon number density is \beq n_b=2.25\times
10^{-4}\ \left(\frac{1+z}{10}\right)^3\ \left(\frac{\Omega_b
h^2}{0.02}\right)\mbox { cm}^{-3}\ .\eeq Throughout the paper we refer
to proper densities rather than comoving densities. As our standard
case we assume a post--reionization ratio of $n_{\gamma}/n_b=1$, but
we also consider the effect of setting $n_{\gamma}/n_b=0.1$. For
example, $\alpha=1.8$ and $n_{\gamma}/n_b=1$ yield $I_{21}=1.0$ at
$z=3$ and $I_{21}=3.5$ at $z=5$, close to the values required to
satisfy the Gunn-Peterson constraint at these redshifts (see, e.g.,
Efstathiou 1992). Note that $n_{\gamma}/n_b\ga 1$ is required for the
initial ionization of the gas in the Universe (although this ratio may
decline after reionization).

We assume that the above uniform UV background illuminates the outer
surface of the gas cloud, located at the virial radius $r_{\rm vir}$,
and penetrates from there into the cloud. The radiation photoionizes
and heats the gas at each radius to its equilibrium temperature,
determined by equating the heating and cooling rates. 
The latter assumption is justified by the fact that both the recombination
time and the heating time are initially shorter than the dynamical
time throughout the halo. At the outskirts of the halo the dynamics
may start to change before the gas can be heated up to its equilibrium
temperature, but this simply means that the gas starts expanding out
of the halo during the process of photoheating. This outflow should
not alter the overall fraction of evaporated gas.

The process of reionization is expected to be highly non-uniform due to the
clustering of the ionizing sources and the clumpiness of the IGM. As time
progresses, the HII regions around the ionizing sources overlap, and each
halo is exposed to ionizing radiation from an ever increasing number of
sources. While the external ionizing radiation may at first be dominated by
a small number of sources, it quickly becomes more isotropic as its
intensity builds up with time (e.g., Haiman \& Loeb 1998a,b; Miralda-Escud\'e,
Haehnelt, \& Rees 1998). The evolution of this process depends on the
characteristic clustering scale of ionizing sources and their
correlation with the inhomogeneities of the IGM. In particular, 
the process takes more time if
the sources are typically embedded in dense regions of the neutral IGM
which need to be ionized first before their radiation shines on the rest of
the IGM. However, in our analysis we do not need to consider these
complications since the total fraction of evaporated gas in bound halos
depends primarily on the maximum intensity achieved at the end of the
reionization epoch.

In computing the effect of the background radiation, we include
self-shielding of the gas which is important at the high densities
obtained in the core of high redshift halos. For this purpose, we
include radiative transfer through the halo gas and photoionization by
the resulting anisotropic radiation field in the calculation of the
ionization equilibrium. 
We also include the fact that the ionizing spectrum becomes harder at
inner radii, since the outer gas layers preferentially block photons
with energies just above the Lyman limit. We neglect self-shielding
due to helium atoms. Appendix B summarizes our simplified treatment of
the radiative transfer equations.

Once the gas is heated throughout the halo, some fraction of it
acquires a sufficiently high temperature that it becomes unbound. This
gas expands due to the resulting pressure gradient and eventually
evaporates back to the IGM. The pressure gradient force (per unit
volume) $k_B \nabla (T \rho/\mu)$ competes with the gravitational
force of $\rho G M/r^2$. Due to the density gradient, the ratio
between the pressure force and the gravitational force is roughly the
ratio between the thermal energy $\sim k_B T$ and the gravitational
binding energy $\sim \mu G M/r$ (which is $\sim k_B T_{\rm vir}$ at
$r_{\rm vir}$) per particle. Thus, if the kinetic energy exceeds the
potential energy (or roughly if $T>T_{\rm vir}$), the repulsive
pressure gradient force exceeds the attractive gravitational force and
expels the gas on a dynamical time (or faster for halos with $T\gg
T_{\rm vir}$).

We compare the thermal and gravitational energy (both of which are
functions of radius) as a benchmark for deciding which gas shells are
expelled from each halo. Note that infall of fresh IGM gas into the
halo is also suppressed due to its excessive gas pressure, produced by
the same photo-ionization heating process.

This situation stands in contrast to feedback due to supernovae, which
depends on the efficiency of converting the mechanical energy of the
supernovae into thermal energy of the halo gas. The ability of supernovae
to disrupt their host dwarf galaxies has been explored in a number of
theoretical papers (e.g., Larson 1974; Dekel \& Silk 1986; Vader 1986,
1987). However, numerical simulations (Mac-Low \& Ferrara 1998) find that
supernovae produce a hole in the gas distribution through which they expel
the shock-heated gas, leaving most of the cooler gas still bound. In the
case of reionization, on the other hand, energy is imparted to the gas
directly by the ionizing photons. A halo for which a large fraction of the
gas is unbound by reionization is thus prevented from further collapse and
star formation.

When the gas in each halo is initially ionized, an ionization shock front
may be generated (cf.\ the discussion of Ly$\alpha$ absorbers by Donahue \&
Shull 1987). The dynamics of such a shock front have been investigated in
the context of the interstellar medium by Bertoldi \& McKee (1990) and
Bertoldi (1989). Their results imply that the dynamics of gas in a halo
are not significantly affected by the shock front unless the thermal energy
of the ionized gas is greater than its gravitational potential energy.
Furthermore, since gas in a halo is heated to the virial temperature even
before reionization, the shock is weaker when the gas is ionized than a
typical shock in the interstellar medium. Also, as noted above, the
ionizing radiation reaching a given halo builds up in intensity over a
considerable period of time. Thus, we do not expect the ionization shock
associated with the first encounter of ionizing radiation to have a large
effect on the eventual fate of gas in the halo.

\section{Results}

We assume the most popular cosmology to date (Garnavich et al.\ 1998) with
$\Omega_0=0.3$ and $\Omega_{\Lambda}=0.7$. We illustrate the effects of
cosmological parameters by displaying the results also for $\Omega_0=1$,
and for $\Omega_0=0.3$ and $\Omega_{\Lambda}=0$. The models all assume
$\Omega_b h^2=0.02$ and a Hubble constant $h=0.5$ if $\Omega_0=1$ and
$h=0.7$ otherwise (where $H_0=100\, h\mbox{ km s}^{-1}\mbox{Mpc}^{-1}$).

Figure 1 shows the temperature of the gas versus its baryonic overdensity
$\Delta_b$ relative to the cosmic average (cf.\ Efstathiou 1992). The
curves are for $z=8$ and assume $\Omega_0=0.3$ and
$\Omega_{\Lambda}=0.7$. We include intergalactic radiation with a flux
given by equation (\ref{Inu}) for $\alpha=1.8$ and $n_{\gamma}/n_b=1$.
The dotted curve shows $t_H=t_{cool}$ with no radiation field, where $t_H$
is the age of the Universe, approximately equal to $6.5\times 10^9 h^{-1}
(1+z)^{-3/2} \Omega_0^{-1/2}$ years at high redshift. This curve indicates
the temperature to which gas has time to cool through atomic transitions before
reionization. This temperature is always near $T=10^4$K since below this
temperature the gas becomes mostly neutral and the cooling time is very
long. It is likely that only atomic cooling is relevant before
reionization since molecular hydrogen is easily destroyed by even a weak
ionizing background (Haiman, Rees, \& Loeb 1996).
The solid curve shows the equilibrium temperature for which the heating time
$t_{heat}$ due to a UV radiation field equals the cooling time $t_{cool}$.
The decrease in the temperature at $\Delta_b<10$ is due to the increased
importance of Compton cooling, which is proportional to the gas density
rather than its square. At a given density, gas is heated at
reionization to the temperature indicated by the solid curve, unless the
net cooling or heating time is too long. The dashed curves show the
temperature where the net cooling or heating time equals $t_H$. By
definition, points on the solid curve have an infinite net cooling or
heating time, but there is also a substantial regime at low $\Delta_b$
where the net cooling or heating time is greater than $t_H$. However, this
regime has only a minor effect on halos, since the mean overdensity inside
the virial radius of a halo is of order 200. On the other hand, if gas
leaves the halo and expands it quickly enters the regime where it cannot
reach thermal equilibrium.

Figure 2 presents an example for the structure of a halo with an
initial total mass of $M=3\times10^7 M_{\sun}$ at $z=8$. We assume
the same cosmological parameters as in Figure 1. The bottom plot shows
the baryon overdensity $\Delta_b$ versus $r/r_{\rm vir}$, and reflects
our assumption of identical NFW profiles for both the dark matter and
the baryons. The middle plot shows the neutral hydrogen fraction
versus $r/r_{\rm vir}$, and the top plot shows the ratio of thermal
energy per particle (${\rm TE}=\frac{3}{2} k_B T$) to potential energy
per particle (${\rm PE}=\mu |\phi(r)|$, where $\phi(r)$ is the
gravitational potential) versus $r/r_{\rm vir}$. The dashed curves
assume an optically thin halo, while the solid curves include
radiative transfer and self-shielding. The self-shielded neutral core
is apparent from the solid curves, but since the point where ${\rm
TE/PE}=1$ occurs outside this core, the overall unbound fraction does
not depend strongly on the radiative transfer in this case. Its value
is $67\%$ assuming an optically--thin halo, and $64\%$ when radiative
transfer is included and only a fraction of the external photons make
their way inside. Even when the opacity at the Lyman limit is large,
some ionizing radiation still reaches the central parts of the halo
because, (i) the opacity drops quickly above the Lyman limit, and (ii)
the heated gas radiates ionizing photons inwards.

Figure 3 shows the unbound gas fraction after reionization as a function of
the total halo mass. We assume $\Omega_0=0.3$, $\Omega_{\Lambda}=0.7$, and
$n_{\gamma}/n_b=1$. The three pairs of curves shown consist of a solid
line (which includes radiative transfer) and a dashed line (which assumes
an optically thin halo). From right to left, the first pair is for
$\alpha=1.8$ and $z=8$, the second is for $\alpha=5$ and $z=8$, and the
third is for $\alpha=1.8$ and $z=20$. In each case the self-shielded core
lowers the unbound fraction when we include radiative transfer (solid vs
dashed lines), particularly when the unbound fraction is sufficiently large
that it includes part of the core itself. High energy photons above the
Lyman limit penetrate deep into the halo and heat the gas
efficiently. Therefore, a steepening of the spectral slope from
$\alpha=1.8$ to $\alpha=5$ decreases the temperature throughout the halo
and lowers the unbound gas fraction. This is only partially compensated for
by our UV flux normalization, which increases $I_{21}$ with increasing
$\alpha$ so as to get the same density of ionizing photons in
equation~(\ref{eq:n_gamma}). Increasing the reionization redshift from
$z=8$ to $z=20$ increases the binding energy of the gas, because the high
redshift halos are denser. Although the corresponding increase of $I_{21}$
with redshift (at a fixed $n_\gamma/n_{b}$) counteracts this change, the
fraction of expelled gas is still reduced due to the deeper
potential wells of higher redshift halos.

 From plots similar to those shown in Figure 3, we find the total halo mass
at which the unbound gas fraction is $50\%$. We henceforth refer to this
mass as the $50\%$ mass. Figure 4 plots this mass as a function of the
reionization redshift for different spectra and cosmological models. The
solid line assumes $\alpha=1.8$ and the dotted line $\alpha=5$, both for
$\Omega_0=0.3$ and $\Omega_{\Lambda}=0.7$. The other lines assume
$\alpha=1.8$ but different cosmologies. The short-dashed line assumes
$\Omega_0=0.3$, $\Omega_{\Lambda}=0$ and the long-dashed line assumes
$\Omega_0=1$. All assume $n_{\gamma}/n_b=1$. Gas becomes unbound when its
thermal energy equals its potential binding energy. The thermal energy
depends on temperature, but the equilibrium temperature does not change
much with redshift since we increase the UV flux normalization by the same
$(1+z)^3$ factor as the mean baryonic density. With this prescription for
the UV flux, the $50\%$ mass occurs at a value of the circular velocity
which is roughly constant with redshift. Thus for each curve, the change in
mass with redshift is mostly due to the change in the characteristic halo
density, which affects the relation between circular velocity and mass.

The cosmological parameters have only a modest effect on the $50\%$ mass,
and change it by up to $35\%$ at a given redshift. Lowering $\Omega_0$
reduces the characteristic density of a halo of given mass, and so a higher
mass is required in order to keep the gas bound. Adding a cosmological
constant reduces the density further through $\Delta_c$ [see
equations~(\ref{dc1}) and (\ref{dc2})]. For the three curves with
$\alpha=1.8$, the circular velocity of the $50\%$ mass equals $13~{\rm
km~s^{-1}}$ at all redshifts, up to variations of a few percent. 

The spectral shape of the ionizing flux affects modestly the threshold
circular velocity corresponding to the $50\%$ mass, because assuming
a steeper spectrum (i.e.\ with a larger $\alpha$) reduces the 
gas temperature and thus requires
a shallower potential to keep the gas bound. A higher flux
normalization has the opposite effect of increasing the threshold circular
velocity. The left panel of Figure 5 shows the variation of circular
velocity with spectral shape, for two normalizations ($n_{\gamma}/n_b=1$
and $n_{\gamma}/n_b=0.1$ for the solid and dashed curves,
respectively). The right panel shows the complementary case of varying the
spectral normalization, using two values for the spectral slope
($\alpha=1.8$ and $\alpha=5$ for the solid and dashed curves,
respectively). All curves assume an $\Omega_0=0.3$, 
$\Omega_{\Lambda}=0.7$ cosmology.

Obviously, $50\%$ is a fairly arbitrary choice for the unbound gas
fraction at which halos evaporate. Figure 3 shows that for a given
halo, the unbound gas fraction changes from $10\%$ to $90\%$ over a
factor of $\sim 60$ in mass, or a factor of $\sim 4$ in velocity
dispersion. When $50\%$ of the gas is unbound, however, the rest of
the gas is also substantially heated, and we expect the process of
collapse and fragmentation to be inhibited. In the extreme case where
the gas expands until a steady state is achieved where it is pressure
confined by the IGM, less than $10\%$ of the original gas is left
inside the virial radius. However, continued infall of dark matter
should limit the expansion. Numerical simulations may be used to
define more precisely the point at which gas halos are disrupted.
Clearly, photo-evaporation affects even halos with masses well above
the $50\%$ mass, although these halos do not completely evaporate.
Note that it is also clear from Figure 3 that not including radiative
transfer would have only a minor effect on the value of the $50\%$
mass (typically $\sim 5\%$).

Given the values of the unbound gas fraction in halos of different masses,
we can integrate to find the total gas fraction in the Universe which
becomes unbound at reionization. This calculation requires the abundance
distribution of halos, which is readily provided by the Press-Schechter
mass function for CDM cosmologies (relevant expressions are given,
e.g., in NFW). The high-mass cutoff in the integration is given by the
lowest mass halo for which the unbound gas fraction is zero, since halos
above this mass are not significantly affected by the UV radiation. The
low-mass cutoff is given by the lowest mass halo in which gas has assembled
by the reionization redshift. We adopt for this low-mass cutoff the linear
Jeans mass, which we calculate following Peebles (1993, \S 6). The gas
temperature in the Universe follows the cosmic microwave background
temperature down to a redshift $1+z_t \sim 740 (\Omega_b h^2)^{2/5}$, at
which the baryonic Jeans mass is $1.9\times 10^5 (\Omega_b h^2)^
{-1/2}M_{\sun}$. After this redshift, the gas temperature goes down as
$(1+z)^2$, so the baryon Jeans mass acquires a factor of
$[(1+z)/(1+z_t)]^{3/2}$. Until now we have considered baryons only,
but if we add dark matter then the mean density 
(or the corresponding gravitational
force) is increased by $\Omega_0/\Omega_b$, which decreases the baryonic
Jeans mass by $(\Omega_0/\Omega_b)^{-3/2}$. The corresponding total halo
mass is $\Omega_0/\Omega_b$ times the baryonic mass. Thus the Jeans cutoff
before reionization corresponds to a total halo mass of \beq M_J=6.9 \times
10^3\ \left(\frac{\Omega_0 h^2}{0.2}\right)^{-\frac{1}{2}}\
\left(\frac{\Omega_b h^2}{0.02}\right)^{-\frac{3}{5}}\
\left(\frac{1+z}{10}\right)^{\frac{3}{2}}\ M_{\sun}\ . \eeq This value
agrees with the numerical spherical collapse calculations of Haiman, Thoul,
\& Loeb (1996).

We thus calculate the total fraction of gas in the Universe which is bound
in pre-existing halos, and the fraction of this gas which then becomes
unbound at reionization. In Figure 6 we show the fraction of the
collapsed gas which evaporates as a function of the reionization redshift.
The solid line assumes $\alpha=1.8$, and the dotted line assumes
$\alpha=5$, both for $\Omega_0=0.3$, $\Omega_{\Lambda}=0.7$. The other
lines assume $\alpha=1.8$, the short-dashed line with $\Omega_0=0.3$,
$\Omega_{\Lambda}=0$ and the long-dashed line with $\Omega_0=1$. All assume
$n_{\gamma}/n_b=1$ and a primordial $n=1$ (scale invariant) power spectrum.
In each case we normalized the CDM power spectrum to the present cluster
abundance, $\sigma_8=0.5\ \Omega_0^{-0.5}$ (see, e.g., Pen 1998), where
$\sigma_8$ is the root-mean-square amplitude of mass fluctuations in
spheres of radius $8\ h^{-1}$ Mpc. The fraction of collapsed gas which is
unbound is $\sim 0.4$--0.7 at $z=6$ and it increases with redshift. This
fraction clearly depends strongly on the halo abundance but is relatively
insensitive to the spectral slope $\alpha$ of the ionizing radiation. In
hierarchical models, the characteristic mass (and binding energy) of
virialized halos is smaller at higher redshifts, and a larger fraction of
the collapsed gas therefore escapes once it is photoheated. Among the three
cosmological models, the characteristic mass at a given redshift is
smallest for $\Omega_0=1$ and largest for $\Omega_0=0.3$,
$\Omega_{\Lambda}=0$.

In Figure 7 we show the total fraction of gas in the Universe which
evaporates at reionization. The solid line assumes $\alpha=1.8$, and the
dotted line assumes $\alpha=5$, both for $\Omega_0=0.3$,
$\Omega_{\Lambda}=0.7$. The other lines assume $\alpha=1.8$, the
short-dashed line with $\Omega_0=0.3$, $\Omega_{\Lambda}=0$ and the
long-dashed line with $\Omega_0=1$. All assume $n_{\gamma}/n_b=1$. For the
different cosmologies, the total unbound fraction goes up to 20--25$\%$ if
reionization occurs as late as $z=6$--$7$; in this case a substantial
fraction of the total gas in the Universe undergoes the process of
expulsion from halos. However, this fraction typically decreases at higher
redshifts. Although a higher fraction of the collapsed gas evaporates at
higher $z$ (see Figure 6), a smaller fraction of the gas in the Universe
lies in halos in the first place. The latter effect dominates except for
the open model up to $z\sim 7$. As is well known, the $\Omega_0=1$ model
produces late structure formation, and indeed the collapsed fraction
decreases rapidly with redshift in this cosmological model. The low
$\Omega_0$ models approach the $\Omega_0=1$ behavior at high $z$, but this
occurs faster for the flat model with a cosmological constant than for the
open model with the same value of $\Omega_0$.

Changing the dark matter and gas profiles as discussed in \S 2 has a modest
effect on the results. For example, with $\Omega_0=0.3$,
$\Omega_{\Lambda}=0.7$, and $z=8$, and for our standard model where the gas
and dark matter follow identical NFW profiles, the total unbound gas
fraction is $19.8\%$ and the halo mass which loses $50\%$ of its baryons is
$5.25\times 10^7 M_{\sun}$. If we let the mass and the baryons follow the
profile of equation (\ref{core}) the corresponding results are $20.0\%$ and
$5.31 \times 10^7 M_{\sun}$ for $b=10$ in equation~(\ref{core}) and
$20.9\%$ and $6.84 \times 10^7 M_{\sun}$ for $b=5$ (i.e.\ a larger core).
With an NFW mass profile but gas in hydrostatic equilibrium at the virial
temperature, the unbound fraction is $19.2\%$, and the $50\%$ mass is
$4.33\times10^7 M_{\sun}$. If we let the gas temperature be $T=2\ T_{\rm
vir}$, the unbound fraction is $22.0\%$ and the $50\%$ mass is $1.18 \times
10^8 M_{\sun}$. For clouds of gas which condense by cooling for a Hubble
time, the unbound fraction is $18.2\%$, and the $50\%$ mass is $3.38\times
10^7 M_{\sun}$. We conclude that centrally concentrated gas clouds are in
general more effective at retaining their gas, but the effect on the
overall unbound gas fraction in the Universe is modest, even for large
variations in the profile. If we return to the NFW profile but adopt
$f=0.01$ instead of $f=0.5$ in the NFW prescription for finding the
collapse redshift (see Appendix A), we find an unbound fraction of
$20.3\%$, and a $50\%$ mass of $6.06\times 10^7 M_{\sun}$. Finally,
lowering $\Omega_b$ by a factor of 2 changes the unbound fraction to
$19.0\%$ and the $50\%$ mass to $5.44\times 10^7 M_{\sun}$. Our predictions
appear to be robust against variations in the model parameters.

\section{Implications for the Intergalactic Medium and for Low
Redshift Objects}

Our calculations show that a substantial fraction of gas in the Universe
may lie in virialized halos before reionization, and that most of it
evaporates out of the halos when it is photoionized and heated at
reionization. The resulting outflows of gas from halos may have interesting
implications for the subsequent evolution of structure in the IGM. We
discuss some of these implications in this section.

In the pre-reionization epoch, a fraction of the gas in the dense cores of
halos may fragment and form stars. Some star formation is, of course,
needed in order to produce the ionizing flux which leads to
reionization. These population III stars produce the first metals in the
Universe, and they may make a substantial contribution to the enrichment of
the IGM. Numerical models by Mac-Low \& Ferrara (1998) suggest that
feedback from supernovae is very efficient at expelling metals from dwarf
galaxies of total mass $3.5\times 10^8\, M_{\sun}$, although it ejects only
a small fraction of the interstellar medium in these hosts. Obviously, the
metal expulsion efficiency depends on the presence of clumps in the
supernova ejecta (Franco et al.\ 1993) and on the supernova rate -- the
latter depending on the unknown star formation rate and the initial mass
function of stars at high redshifts. Reionization provides an alternative
method for expelling metals efficiently out of dwarf galaxies by directly
photoheating the gas in their halos, leading to its evaporation along with
its metal content.\footnote{Note that we have assumed zero metallicity in
calculating cooling. Even if some metals had already been mixed into the
IGM, the metallicity of newly formed objects 
was likely too low to affect cooling since even at
$z\sim 3$ the typical metallicity of the Lyman alpha forest has been
observed to be $<0.01$ solar (Songaila \& Cowie 1996; Tytler et al.\ 
1995).}

Gas which falls into halos and is expelled at reionization attains a
different entropy than if it had stayed at the mean density of the
Universe. Gas which collapses into a halo is at a high overdensity
when it is photoheated, and is therefore at a lower entropy than if it
were heated to the same temperature at the mean cosmic density.
However, the overall change in the entropy density of the IGM is small
for two reasons. First, even at $z=6$ only about $25\%$ of the gas in
the Universe undergoes evaporation. Second, the gas remains in
ionization equilibrium and is photoheated during its initial
expansion. For example, if $z=6$, $\Omega_0=0.3$,
$\Omega_\Lambda=0.7$, $n_{\gamma}/n_b=1$, and $\alpha=1.8$, then the
recombination time becomes longer than the dynamical time only when
the gas expands down to an overdensity of 26, at which point its
temperature is 22,400 K compared to an initial (non-equilibrium)
temperature of 19,900 K for gas at the mean density. The resulting
overall reduction in the entropy is the same as would be produced by
reducing the temperature of the entire IGM by a factor of 1.6. This
factor reduces to 1.4 if we increase $z$ to 8 or increase $\alpha$ to
5. Note that Haehnelt \& Steinmetz (1998) showed that differences in
temperature by a factor of $3$--$4$ result in possibly observable
differences in the Doppler parameter distribution of Ly$\alpha$
absorption lines at redshifts 3--5.

When the halos evaporate, recombinations in the gas could produce
Ly$\alpha$ lines or radiation from two-photon transitions to the
ground state of hydrogen. However, a simple estimate shows that the
resulting luminosity is too small for direct detection unless these
halos are illuminated by an internal ionizing source. In an externally
illuminated $z=6$, $10^8 M_{\sun}$ halo our calculations imply a total
of $\sim 1\times 10^{50}$ recombinations per second. Note that the
number of recombinations is dominated by the high density core, and if
we did not include self-shielding we would obtain an overestimate by a
factor of $\sim 15$. If each recombination releases one or two photons
with a total energy of $10.2$ eV, then for $\Omega_0=0.3$ and
$\Omega_\Lambda=0.7$ the observed flux is $\sim 5\times 10^{-20}$ erg
s$^{-1}$ cm$^{-2}$. This flux is well below the sensitivity of the
planned Next Generation Space Telescope, even if part of this flux is
concentrated in a narrow line.

The photoionization heating of the gaseous halos of dwarf galaxies resulted
in outflows with a characteristic velocity of $\sim 20$--$30~{\rm
km~s^{-1}}$. These outflows must have induced peculiar velocities of a
comparable magnitude in the IGM surrounding these galaxies. 
The effect of the outflows on the velocity field and entropy of the 
IGM at $z=5$--10 could in principle be searched for in the absorption 
spectra of high redshift sources, such as quasars. 
These small-scale fluctuations in velocity and the resulting temperature
fluctuations have been seen in recent simulations by Bryan et al.\ (1998).
However, the small halos responsible for these outflows were only 
barely resolved even in these high resolution simulations of a small volume.

The evaporating galaxies could contribute to the high column density
end of the Ly$\alpha$ forest (cf.\ Bond, Szalay, \& Silk 1988). For
example, shortly after being photoionized, a $z=8$, $5\times 10^7\
M_{\sun}$ halo has a neutral hydrogen column density of $2\times
10^{16}$ cm$^{-2}$ at an impact parameter of $0.5\, r_{\rm vir}=0.66$
kpc, $6\times 10^{17}$ cm$^{-2}$ at $0.25\, r_{\rm vir}$, and $9\times
10^{20}$ cm$^{-2}$ (or $9\times 10^{18}$ cm$^{-2}$ if we do not
include self-shielding) at $0.1\, r_{\rm vir}$ (assuming
$\Omega_0=0.3$, $\Omega_{\Lambda}=0.7$, $\alpha=1.8$, and
$n_{\gamma}/n_b=1$). These column densities will decline as the gas
expands out of the host galaxy. Abel \& Mo (1998) have suggested that
a large fraction of the Lyman limit systems at $z\sim 3$ may
correspond to mini-halos that survived reionization. Remnant
absorbers due to galactic outflows can be distinguished from
large-scale absorbers in the IGM by their compactness. Close lines of
sight due to quasar pairs or gravitational lensed quasars (see, e.g.,
Crotts \& Fang 1998; Petry, Impey, \& Foltz 1998, and references
therein) should probe different HI column densities in galactic
outflow absorbers but similar column densities in the larger, more
common absorbers. Follow-up observations with high spectroscopic
resolution could reveal the velocity fields of these outflows.

%The Jeans mass after reionization is $\sim 10^10 M_{\sun}$, which
%corresponds to $V_c \sim 80\ {\rm km\ s}^{-1}$ for a virialized halo of
%this mass. However, the dark matter is unaffected by pressure and
%it collapses also at smaller mass scales. The resulting high 
%overdensities allow gas to fall onto smaller halos, although the
%gas fraction is reduced in halos below the Jeans mass.

Although much of the gas in the Universe evaporated at reionization,
the underlying dark matter halos continued to evolve through infall
and merging, and the heated gas may have accumulated in these halos at
lower redshifts. This latter process has been discussed by a number
of authors, with an emphasis on the effect of reionization and the
resulting heating of gas. Thoul \& Weinberg (1996) found a reduction
of $\sim50\%$ in the collapsed gas mass due to heating, for a halo of
$V_c=50\ {\rm km\ s}^{-1}$ at $z=2$, and a complete suppression of
infall below $V_c=30\ {\rm km\ s}^{-1}$. The effect is thus
substantial on halos with virial temperatures well above the gas
temperature. Their interpretation is that pressure support delays
turnaround substantially and slows the subsequent collapse. Indeed, as
noted in \S 2, the ratio of the pressure force to the gravitational
force on the gas is roughly equal to the ratio of its thermal energy
to its potential energy. For a given enclosed mass, the potential
energy of a shell of gas increases as its radius decreases. Before
collapse, each gas shell expands with the Hubble flow until its
expansion is halted and then reversed. Near turnaround, the gas is
weakly bound and the pressure gradient may prevent collapse even for
gas below the halo virial temperature. On the other hand, gas which
is already well within the virial radius is tightly bound, which
explains our lower value of $V_c \sim 13\ {\rm km\ s}^{-1}$ for halos
which lose half their gas at reionization.

Three dimensional numerical simulations (Quinn, Katz, \& Efstathiou
1996; Weinberg, Hernquist, \& Katz 1997; Navarro \& Steinmetz 1997)
have also explored the question of whether dwarf galaxies could
re-form at $z \ga 2$. The heating by the UV background was found to
suppress infall of gas into even larger halos ($V_c \sim 75\ {\rm km\
s}^{-1}$), depending on the redshift and on the ionizing radiation
intensity. Navarro \& Steinmetz (1997) noted that photoionization
reduces the cooling efficiency of gas at low densities, which
suppresses further the late infall at redshifts below 2. We note that
these various simulations assume an isotropic ionizing radiation
field, and do not calculate radiative transfer. Photoevaporation of a
gas cloud has been calculated in a two dimensional simulation
(Shapiro, Raga, \& Mellema 1998), and methods are being developed for
incorporating radiative transfer into three dimensional cosmological
simulations (e.g., Abel, Norman, \& Madau 1999; Razoumov \& Scott
1999).

Our results have interesting implications for the fate of gas in
low-mass halos. Gas evaporates at reionization from halos below $V_c
\sim 13\ {\rm km\ s}^ {-1}$, or a velocity dispersion $\sigma \sim 10\
{\rm km\ s}^ {-1}$. A similar value of the velocity dispersion is also
required to reach a virial temperature of $10^4$ K, allowing atomic
cooling and perhaps star formation before reionization. Thus, halos
with $\sigma \ga 10\ {\rm km\ s}^{-1}$ could have formed stars before
reionization. They would have kept their gas after reionization, and
could have had ongoing star formation subsequently. These halos were
the likely sites of population III stars, and could have been the
progenitors of dwarf galaxies in the local Universe (cf.\
Miralda-Escud\'e \& Rees 1998). On the other hand, halos with $\sigma
\la 10\ {\rm km\ s}^{-1}$ could not have cooled before
reionization. Their warm gas was completely evaporated from them at
reionization, and could not have returned to them until very low
redshifts, possibly $z\la 1$, so that their stellar population should
be relatively young.

It is interesting to compare these predictions to the properties of
dwarf spheroidal galaxies in the Local Group which have low central
velocity dispersions. At first sight this appears to be a difficult
task. The dwarf galaxies vary greatly in their properties, with many
showing evidence for multiple episodes of star formation as well as
some very old stars (see the recent review by Mateo 1998). Another
obstacle is the low temporal resolution of age indicators for old
stellar populations. For example, if $\Omega_0=0.3$ and
$\Omega_{\Lambda}=0.7$ then the age of the Universe is $43\%$ of its
present age at $z=1$ and $31\%$ at $z=1.5$. Thus, stars that formed at
these redshifts may already be $\sim 10$ Gyr old at present, and are
difficult to distinguish from stars that formed at $z > 5$.

Nevertheless, one of our robust predictions is that most early halos with
$\sigma \la 10\ {\rm km\ s}^{-1}$ could not have formed stars in the
standard hierarchical scenario. Globular clusters belong to one class of
objects with such a low velocity dispersion. Peebles \& Dicke (1968)
originally suggested that globular clusters may have formed at high
redshifts, before their parent galaxies. However, in current cosmological
models, most mass fluctuations on globular cluster scales were unable to
cool effectively and fragment until $z\sim 10$, and were evaporated
subsequently by reionization. We note that Fall \& Rees (1985) proposed an
alternative formation scenario for globular clusters involving a thermal
instability inside galaxies with properties similar to those of the Milky
Way. Globular clusters have also been observed to form in galaxy mergers
(e.g., Miller et al.\ 1997). It is still possible that some of the very
oldest and most metal poor globular clusters originated from $z\ga 10$,
before the UV background had become strong enough to destroy the molecular
hydrogen in them. However, primeval globular clusters should have retained
their dark halos but observations suggest that globular clusters are not
embedded in dark matter halos (Moore 1996; Heggie \& Hut 1995).

Another related population is the nine dwarf spheroidals in the Local
Group with central velocity dispersions $\sigma \la 10\ {\rm km\
s}^{-1}$, including five below $7\ {\rm km\ s}^{-1}$ (e.g., Mateo
1998). In the hierarchical clustering scenario, the dark matter in a
present halo was most probably divided at reionization among several
progenitors which have since merged. The velocity dispersions of these
progenitors were likely even lower than that of the final halo. Thus
the dwarf galaxies could not have formed stars at high redshifts, and
their formation presents an intriguing puzzle. There are two possible
solutions to this puzzle, (i) the ionizing background dropped
dramatically at low redshifts, allowing the dwarf galaxies to form at
$z\la 1$, or (ii) the measured stellar velocity dispersions of the
dwarf galaxies are well below the velocity dispersions of their dark
matter halos.

Unlike globular clusters, the dwarf spheroidal galaxies are
dark matter dominated. The dark halo of a present-day dwarf galaxy may
have virialized at high redshifts but accreted its gas at low redshift
from the IGM. However, for dark matter halos accumulating primordial gas,
Kepner, Babul, \& Spergel (1997) found that even if $I_{21}(z)$
declines as $(1+z)^4$ below $z=3$, only halos with
$V_c \ga 20\ {\rm km\ s}^{-1}$ can form atomic hydrogen by $z=1$, and $V_c
\ga 25\ {\rm km\ s}^{-1}$ is required to form molecular hydrogen.

Alternatively, the dwarf dark halos could have accreted cold gas at
low redshift from a larger host galaxy rather than from the IGM. As
long as the dwarf halos join their host galaxy at a redshift much
lower than their formation redshift, they will survive disruption due
to their high densities. The subsequent accretion of gas could result
from passages of the dwarf halos through the gaseous tidal tail of a
merger event or through the disk of the parent galaxy. In this case,
retainment of cold, dense, and possibly metal enriched gas against
heating by the UV background requires a shallower potential well than
accumulating warm gas from the IGM. Simulations of galaxy encounters
(Barnes \& Hernquist 1992; Elmegreen, Kaufman, \& Thomasson 1993) have
found that dwarf galaxies could form but with small amounts of dark
matter. However, the initial conditions of these simulations assumed
parent galaxies with a smooth dark matter distribution rather than
clumpy halos with dense sub-halos inside them. Simulations by Klypin
et al.\ (1999) suggest that galaxy halos may have large numbers of
dark matter satellites, most of which have no associated stars. If
true, this implies that the dwarf spheroidal galaxies might be
explained even if only a small fraction of dwarf dark halos accreted gas
and formed stars.

A common origin for the Milky Way's dwarf satellites (and a number of
halo globular clusters), as remnants of larger galaxies accreted by
the Milky Way galaxy, has been suggested on independent grounds. These
satellites appear to lie along two (e.g., Majewski 1994) or more
(Lynden-Bell \& Lynden-Bell 1995, Fusi-Pecci et al.\ 1995) polar great
circles. The star formation history of the dwarf galaxies (e.g.,
Grebel 1998) constrains their merger history, and implies that the
fragmentation responsible for their appearance must have occured early
in order to be consistent with the variation in stellar populations
among the supposed fragments (Unavane, Wyse, \& Gilmore 1996;
Olszewski 1998). Observations of interacting galaxies (outside the
Local Group) also suggest the formation of ``tidal dwarf galaxies''
(e.g., Duc \& Mirabel 1997).

Finally, there exists the possibility that the measured velocity dispersion
of stars in the dwarf spheroidals underestimates the velocity dispersion of
their dark halos. Assuming that the stars are in equilibrium, their
velocity dispersion could be lower than that of the halo if the mass
profile is shallower than isothermal beyond the stellar core radius. As
discussed in \S 2, halo profiles are thought to vary from being shallow in
a central core to being steeper than isothermal at larger distances. 
The velocity dispersion and mass to light ratio of a dwarf spheroidal
could also appear high if it is non-spherical or the stellar orbits
are anisotropic. Some dwarf spheroidals may even not be dark matter 
dominated if they are tidally disrupted (e.g., Kroupa 1997).
The observed properties of dwarf spheroidals require a central mass
density of order $0.1 M_{\sun}$ pc$^{-3}$ (e.g., Mateo 1998), which is
$\sim 7\times 10^5$ times the present critical density. The stars therefore
reside either in high-redshift halos or in the very central parts of low
redshift halos. Detailed observations of the velocity dispersion profiles
of these stars could be used to discriminate between these possibilities.

\section{Conclusions}

We have shown that the photoionizing background radiation which filled the
Universe during reionization likely boiled most of the virialized gas
out of CDM halos at that time. The evaporation process probably lasted of
order a Hubble time due to the gradual increase in the UV background as the
HII regions around individual sources overlapped and percolated until the
radiation field inside them grew up to its cosmic value -- amounting to the
full contribution of sources from the entire Hubble volume. The precise
reionization history depends on the unknown star formation efficiency and
the potential existence of mini-quasars in newly formed halos (Haiman \&
Loeb 1998a).

The total fraction of the cosmic baryons which participate in the
evaporation process depends on the reionization redshift, the ionizing
intensity, and the cosmological parameters, but is not very sensitive to
the precise gas and dark matter profiles of the halos. The central core of
halos is typically shielded from the external ionizing radiation by the
surrounding gas, but this core typically contains $<20\%$ of the halo gas
and has only a weak effect on the global behavior of the gas. We have
found that halos are disrupted up to a circular velocity $V_c \sim 13\ {\rm
km\ s}^{-1}$ for a shallow, quasar-like spectrum, or $V_c \sim 11\ {\rm
km\ s}^{-1}$ for a stellar spectrum, assuming the photoionizing sources
build up a density of ionizing photons comparable to the mean cosmological
density of baryons. At this photoionizing intensity, the value of the
circular velocity threshold is nearly independent of redshift. The
corresponding halo mass changes, however, from $\sim 10^8 M_{\sun}$
at $z=5$ to $\sim 10^7 M_{\sun}$ at $z=20$, assuming a shallow ionizing
spectrum.

Based on these findings, we expect that both globular clusters and Local
Group dwarf galaxies with velocity dispersions $\la 10~{\rm km~s^{-1}}$
formed at low redshift, most probably inside larger galaxies. The latter
possibility has been suggested previously for the Milky Way's dwarf 
satellites based on their location along polar great circles.

\acknowledgements

We are grateful to Jordi Miralda-Escud\'{e}, Chris McKee, Roger Blandford, 
Lars Hernquist, and David Spergel for useful discussions. We also thank
Renyue Cen and Jordi Miralda-Escud\'{e} for assistance with the reaction
and cooling rates. RB acknowledges support from Institute Funds. This work
was supported in part by the NASA NAG 5-7039 grant (for AL).

\bigskip
\centerline{\bf APPENDIX A: Halo profile}

We follow the prescription of NFW for obtaining the density profiles of
dark matter halos, but instead of adopting a constant overdensity of 200 we
use the fitting formula of Bryan \& Norman (1998) for the virial
overdensity: 
\beq 
\Delta_c=18\pi^2+82 d-39 d^2 \label{dc1} 
\eeq 
for a flat
Universe with a cosmological constant and 
\beq \Delta_c=18\pi^2+60 d-32 d^2
\label{dc2} 
\eeq 
for an open Universe, where $d\equiv \Omega(z)-1$. Given
$\Omega_0$ and $\Omega_{\Lambda}$, we define 
\beq \Omega(z)=\frac{\Omega_0
(1+z)^3}{\Omega_0 (1+z)^3+\Omega_{\Lambda}+(1-\Omega_0-\Omega_{\Lambda})
(1+z)^2}\ . 
\eeq

In equation (\ref{NFW}) $c$ is determined for a given $\delta_c$ by the
relation \beq \delta_c=\frac{\Delta_c}{3}\frac{c^3}{\ln(1+c)-c/(1+c)} \
. \eeq 
The characteristic density is given by \beq \delta_c=C(f)
\Omega(z)\left(\frac{1+z_{coll}} {1+z}\right)^3\ . \eeq 
For a given halo of
mass $M$, the collapse redshift $z_{coll}$ is defined as the time at which
a mass $M/2$ was first contained in progenitors more massive than some
fraction $f$ of $M$. This is computed using the extended Press-Schechter
formalism (e.g.\ Lacey \& Cole 1993). NFW find that $f=0.01$ fits their
$z=0$ simulation results best. Since we are interested in high redshifts
when mergers are very frequent, we adopt the more natural $f=0.5$ but also
check the $f=0.01$ case.
[For example, the survival time
of a $z=8$, $5\times 10^7\ M_{\sun}$ halo before it merges is $\sim
30$--$40\%$ of the age of the Universe at that redshift (Lacey \& Cole 1993).]
In both cases we adopt the normalization of NFW, which is $C(0.5)=2 \times
10^4$ and $C(0.01)= 3 \times 10^3$.

\bigskip
\centerline{\bf APPENDIX B: Radiative Transfer}

We neglect atomic transitions of helium atoms in the radiative transfer
calculation. We only consider halos for which $k_B T$ is well below
the ionization energy of hydrogen, and so following Tajiri \& Umemura
(1998) we assume that recombinations to excited levels do not result in
further ionizations. On the other hand, recombinations to the ground state
result in the emission of ionizing photons all of which are in a narrow
frequency band just above the Lyman limit frequency $\nu=\nu_L$. 
We follow separately these emitted
photons and the external incoming radiation. The external photons undergo
absorption with an optical depth at the Lyman limit determined by 
\beq \frac{d\tau_{\nu_L}}{ds}=\sigma_{HI}(\nu_L) n_{HI}\ . \eeq 
The emitted photons near $\nu_L$ are propagated by the equation of 
radiative transfer, \beq \frac{dI_{\nu}}{ds}=-\sigma_{HI}(\nu) 
n_{HI}I_{\nu}+\eta_{\nu}\ . \eeq 
Assuming all emitted photons are just above $\nu=\nu_L$, we can set
$\sigma_{HI}(\nu)=\sigma_{HI}(\nu_L)$ in this equation and propagate the
total number flux of ionizing photons, \beq F_1\equiv\int_{\nu_L}^{\infty}
\frac{I_{\nu}}{h \nu}d\nu\ . \eeq The emissivity term for this quantity is
\beq\int_{\nu_L}^{\infty}\frac{\eta_{\nu}}{h \nu}d\nu= \frac{\omega}{4\pi}
\alpha_{HI} n_{HII}n_e \ , \eeq where $\alpha_{HI}$ is the total
recombination coefficient to all bound levels of hydrogen and $\omega$ is
the fraction of recombinations to the ground state. In terms of Table 5.2
of Spitzer (1978), $\omega=(\phi_1-\phi_2)/\phi_1$. We find that a
convenient fitting formula up to $64,000\ $K, accurate to $2\%$, is (with
$T$ in K) \beq \omega=0.205-0.0266\ln(T)+0.0049 \ln^2(T)\ . \eeq

When these photons are emitted they carry away the kinetic energy of the
absorbed electron. When the photons are re-absorbed at some distance from
where they were emitted, they heat the gas with this extra energy.
Since $k_B T \ll h \nu_L$ we do not need to compute the exact
frequency distribution of these photons. Instead we solve a single
radiative transfer equation for the total flux of energy (above the
ionization energy of hydrogen)
in these photons, \beq F_2\equiv\int_{\nu_L}^{\infty} \frac{I_{\nu}}{h
\nu}(h \nu-h \nu_L)d\nu\ . \eeq The emissivity term for radiative transfer
of $F_2$ is \beq \int_{\nu_L}^{\infty}\frac{\eta_{\nu}}{h \nu}(h \nu- h
\nu_L)d\nu=\frac{2.07\times 10^{-11}}{T^{1/2}}\, \frac{
\chi_1(\beta)-\chi_2(\beta)} {4\pi}\, n_{HII}n_e
\mbox{ erg cm}^{-3}\mbox{ s}^{-1}\mbox{ sr}^{-1}\ ,
\eeq where $\beta=h \nu_L/k T$, $T$ is in K, and
the functions $\chi_1$ and $\chi_2$ are given in Table 6.2 of Spitzer
(1978). We find a fitting formula up to $64,000\ $K, accurate to $2\%$
(with $T$ in K):
\beq \chi_1(T)-\chi_2(T)=\left\{ \begin{array}{ll} 0.78 & \mbox{if
$T<10^3$ K} \\ -0.172+0.255\ln(T)-0.0171\ln^2(T) & \mbox{otherwise.}
\end{array} \right. \eeq

  From each point we integrate along all lines of sight to find
$\tau_{\nu_L}$, $F_1$ and $F_2$ as a function of angle. Because of
spherical symmetry, we do this only at each radius, and the angular
dependence only involves $\theta$, the angle relative to the radial
direction. We then integrate to find the photoionization rate. For each
atomic species, the rate is \beq \Gamma_{\gamma i}= \int_0^{4\pi}d\Omega
\int_{\nu_i}^{\infty} \frac{I_{\nu}}{h \nu} \sigma_i(\nu) d\nu\ {\rm
s}^{-1}\ , \eeq where $\nu_i$ and $\sigma_i(\nu)$ are the threshold
frequency and cross section for photoionization of species $i$, given in
Osterbrock [1989; see Eq. (2.31) for HI, HeI and HeII]. For the external
photons the UV intensity is $I_{\nu,0}e^{-\tau_{\nu}}$, with the boundary
intensity $I_{\nu,0}=I_{\nu_L,0} (\nu/\nu_L)^{-\alpha}$ as before, and
$\tau_{\nu}$ approximated as $\tau_{\nu_L} (\nu/\nu_L)^{-3}$. Since
$\sigma_i(\nu)$ has the simple form of a sum of two power laws, the
frequency integral in $\Gamma_ {\gamma i}$ can be done analytically, and
only the angular integration is computed numerically (cf.\ the similar but
simpler calculation of Tajiri \& Umemura 1998). There is an additional
contribution to photoionization for HI only, from the emitted photons just
above $\nu=\nu_L$, given by $\int_0^{4\pi}d\Omega\ \sigma_i(\nu_L) F_1$.
The photoheating rate per unit volume is $n_i \epsilon_i$, where $n_i$ is
the number density of species $i$ and \beq \epsilon_i= \int_0^{4\pi}d\Omega
\int_{\nu_i}^{\infty} \frac{I_{\nu}}{h \nu} \sigma_i(\nu) (h \nu-h \nu_L)
d\nu\ {\rm s}^{-1} {\rm ergs\ s^{-1}}\ . \eeq The rate for the external UV
radiation is calculated for each atomic species similarly to the
calculation of $\Gamma_{\gamma i}$. The emitted photons contribute to
$\epsilon_{HI}$ an extra amount of $\int_0^{4\pi}d\Omega\ \sigma_i(\nu_L)
F_2$.

%%%%%%%% Figure 1
\begin{figure}
\epsscale{0.7}
%\plotone{avi12b.ps}
\plotone{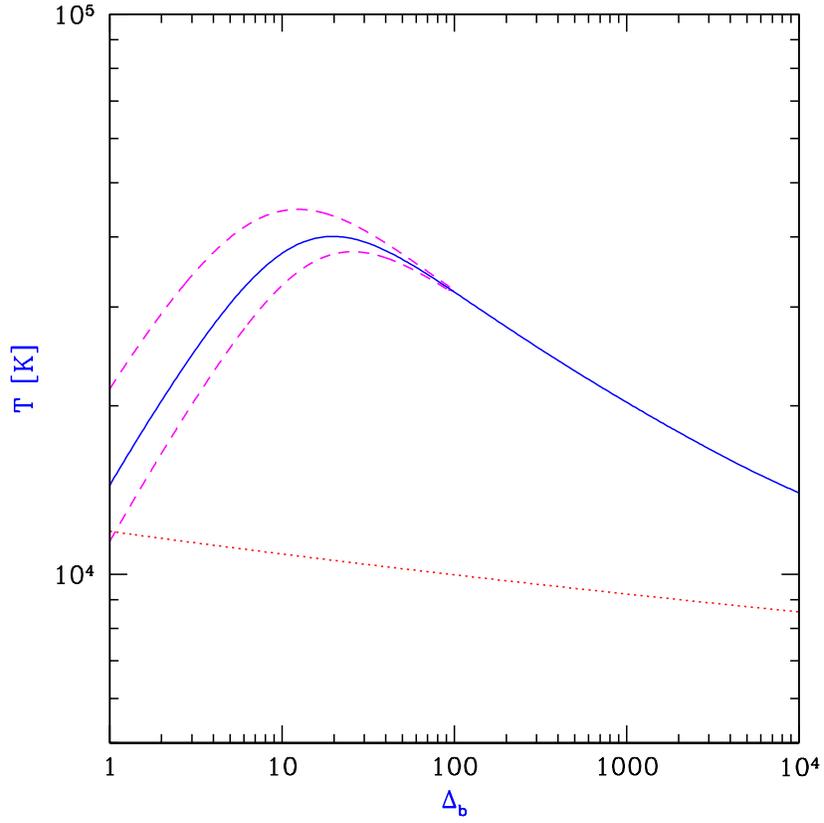}
\caption{Temperature $T$
versus baryon overdensity $\Delta_b$. Dotted curve: 
$t_H=t_{cool}$, no radiation field. Solid curve: 
$t_{heat}=t_{cool}$, with radiation. Dashed curves:
$t_H=$net cooling/heating time, with radiation.
The curves assume $\Omega_0=0.3$, $\Omega_{\Lambda}=0.7$,
$z=8$, $\alpha=1.8$ and $n_{\gamma}/n_b=1$.}
\end{figure}
%%%%%%%%

%%%%%%%% Figure 2
\begin{figure}
\epsscale{0.7}
%\plotone{avi7b.ps}
\plotone{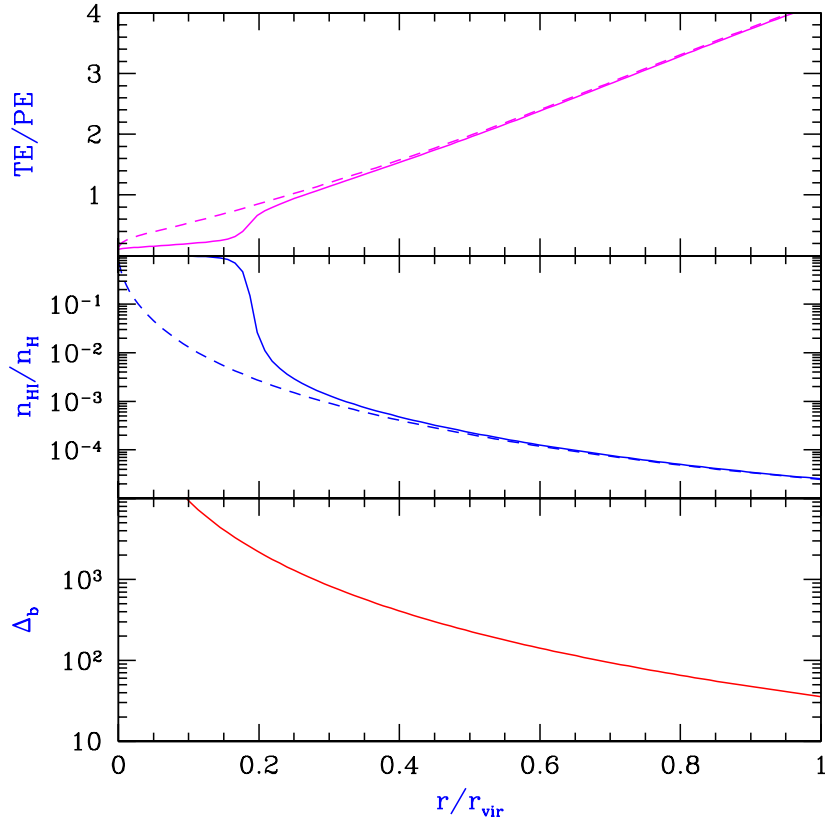}
\caption{Halo structure after reionization, for a halo mass $M=3\times10^7
M_{\sun}$ at $z=8$. Bottom plot: Baryon overdensity ($\Delta_b$) versus
$r/r_{\rm vir}$. Middle plot: Neutral hydrogen fraction ($n_{HI}/n_H$) versus
$r/r_{\rm vir}$. Top plot: Ratio of thermal energy (${\rm TE}$) to
potential energy (${\rm PE}$) versus $r/r_{\rm vir}$. The dashed curves
assume an optically thin halo, while the solid curves include radiative
transfer and self-shielding. All the curves assume $\Omega_0=0.3$,
$\Omega_{\Lambda}=0.7$, $\alpha=1.8$ and $n_{\gamma}/n_b=1$.}
\end{figure}
%%%%%%%%

%%%%%%%% Figure 3
\begin{figure}
\epsscale{0.7}
%\plotone{avi22b.ps}
\plotone{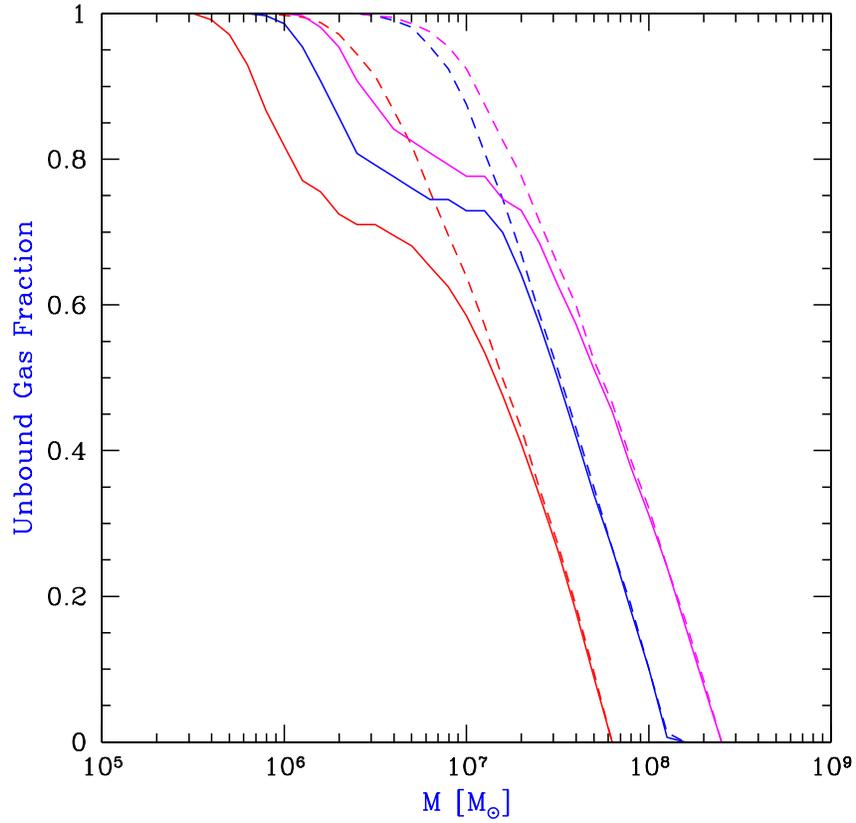}
\caption{Unbound gas fraction versus total halo 
mass. There are three pairs of curves, each consisting of
a solid line (with radiative transfer) and a dashed
line (without radiative transfer). From right to left,
the three sets of curves correspond to
$\alpha=1.8$, $z=8$; $\alpha=5$, $z=8$; and
$\alpha=1.8$, $z=20$. All the curves assume $\Omega_0=0.3$,
$\Omega_{\Lambda}=0.7$, and $n_{\gamma}/n_b=1$.}
\end{figure}
%%%%%%%%

%%%%%%%% Figure 4
\begin{figure}
\epsscale{0.7}
%\plotone{avi30b.ps}
\plotone{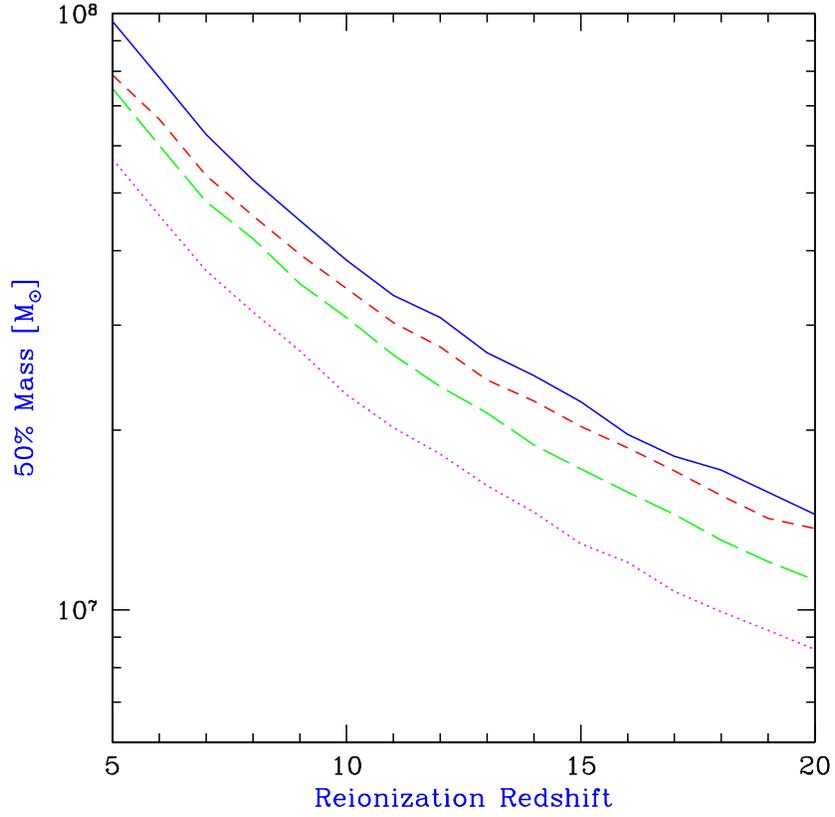}
\caption{Total halo mass for which 50\% of the gas is unbound versus
reionization redshift. The solid line assumes $\alpha=1.8$, and the dotted
line assumes $\alpha=5$, both for an $\Omega_0=0.3$, $\Omega_{\Lambda}=0.7$
cosmology. The other lines assume $\alpha=1.8$ but different
cosmologies. The short-dashed line assumes $\Omega_0=0.3$,
$\Omega_{\Lambda}=0$ and the long-dashed line assumes $\Omega_0=1$. All
assume $n_{\gamma}/n_b=1$.}
\end{figure}
%%%%%%%%

%%%%%%%% Figure 5
\begin{figure}
\epsscale{0.7}
%\plotone{avi34b.ps}
\plotone{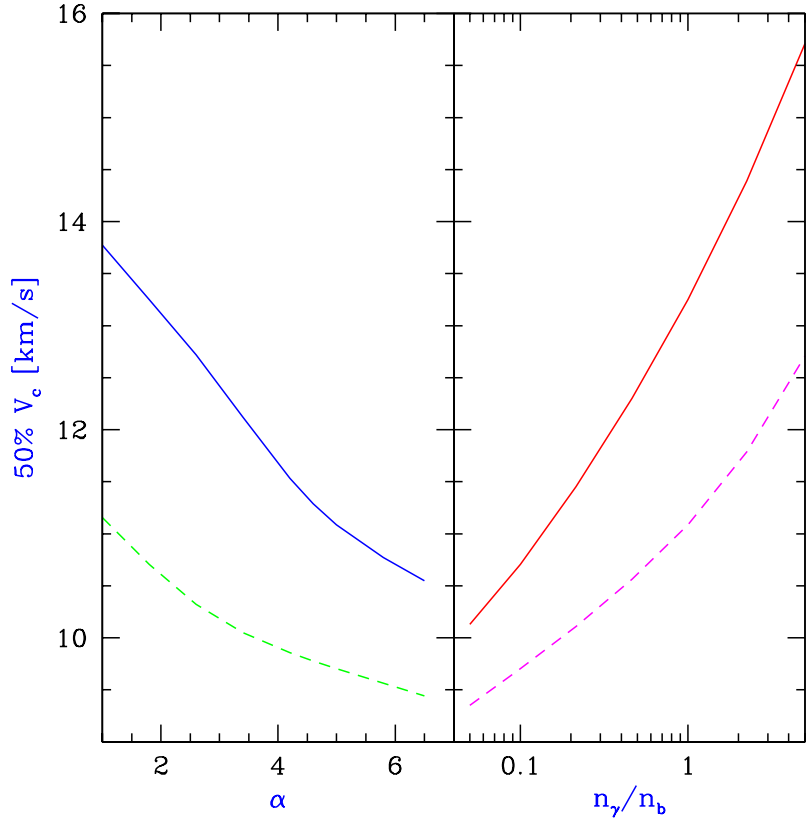}
\caption{Circular halo velocity at which 50\% of the gas is unbound, as a
function of ionizing spectrum. Both panels show $V_c$ on the vertical
axis. The left panel varies the spectral slope $\alpha$ for two values of
the normalization, $n_{\gamma}/n_b=1$ (solid curve) and
$n_{\gamma}/n_b=0.1$ (dashed curve). The right panel varies the
normalization for two spectral slopes, $\alpha=1.8$ (solid curve) and
$\alpha=5$ (dashed curve). All curves assume an $\Omega_0=0.3$, 
$\Omega_{\Lambda}=0.7$ cosmology.}
\end{figure}
%%%%%%%%

%%%%%%%% Figure 6
\begin{figure}
\epsscale{0.7}
%\plotone{avi44.ps}
\plotone{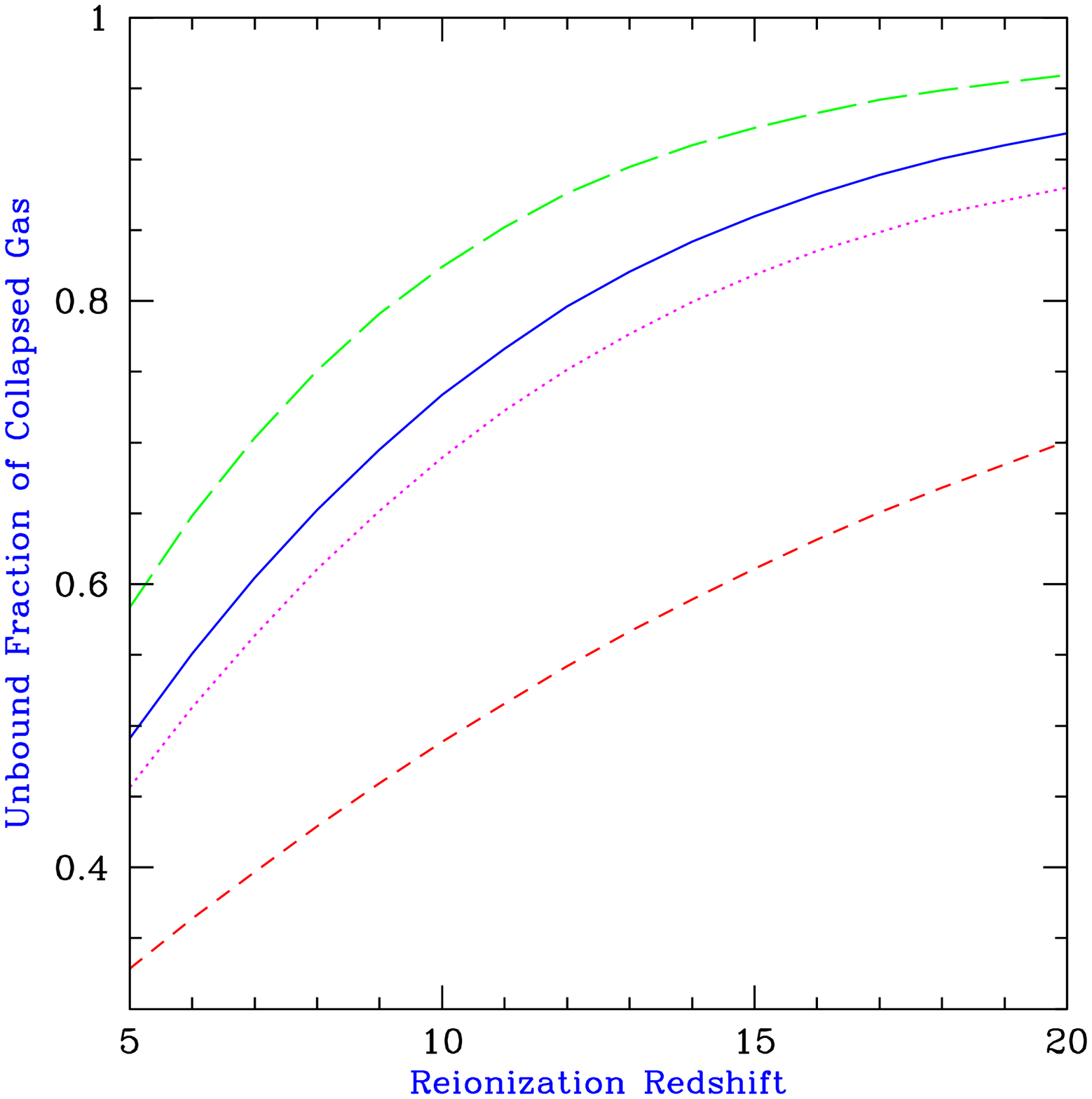}
\caption{Fraction of the collapsed gas which
evaporates from halos at reionization,
versus the reionization redshift.
The solid line assumes $\alpha=1.8$, and the dotted line assumes 
$\alpha=5$, both for an $\Omega_0=0.3$, $\Omega_{\Lambda}=0.7$
cosmology. The other lines assume $\alpha=1.8$ but different 
cosmologies. The short-dashed line
assumes $\Omega_0=0.3$, $\Omega_{\Lambda}=0$ and the long-dashed 
line assumes $\Omega_0=1$. All assume $n_{\gamma}/n_b=1$.}
\end{figure}
%%%%%%%%

%%%%%%%% Figure 7
\begin{figure}
\epsscale{0.7}
%\plotone{avi31.ps}
\plotone{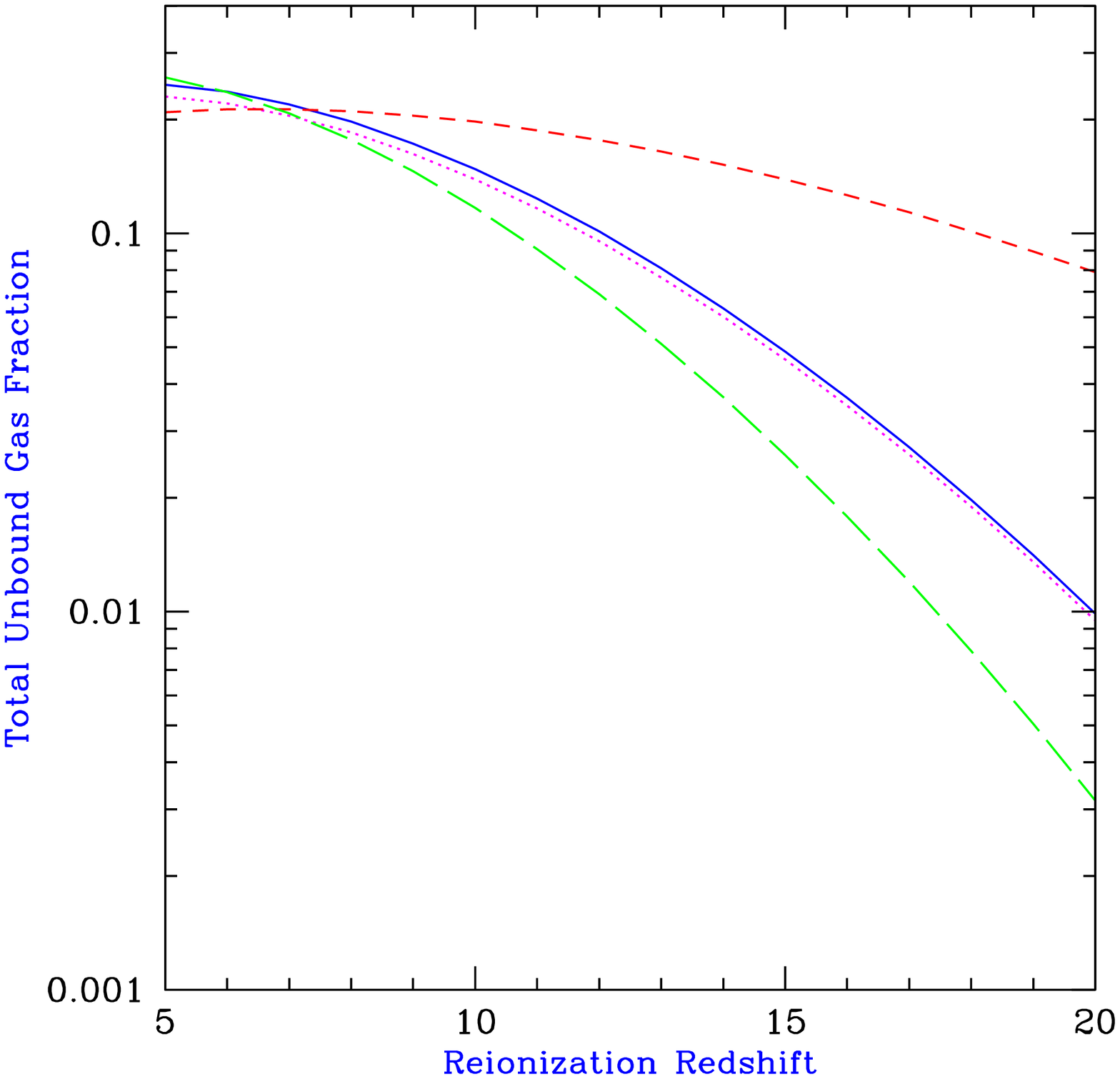}
\caption{Total fraction of gas in the Universe which evaporates from halos
at reionization, versus the reionization redshift. The solid line assumes
$\alpha=1.8$, and the dotted line assumes $\alpha=5$, both for an
$\Omega_0=0.3$, $\Omega_{\Lambda}=0.7$ cosmology. The other lines assume
$\alpha=1.8$ but different cosmologies. The short-dashed line assumes
$\Omega_0=0.3$, $\Omega_{\Lambda}=0$ and the long-dashed line assumes
$\Omega_0=1$. All assume $n_{\gamma}/n_b=1$.}
\end{figure}
%%%%%%%%

\end{document}